\documentclass[times, 10pt,twocolumn]{article} 
\usepackage{times,latex8,epsfig,graphicx,amssymb,latexsym,lscape}

\newcommand{\mr}{\hbox{{\it merge}}}
\newcommand{\spl}{\hbox{{\it split}}}

\newcommand{\cpp}{\hbox{{\tt C++}}}

\begin{document}

\title{Object-Oriented Modeling of Programming Paradigms}

\author{
	M.H. van Emden\\
    University of Victoria\\
    Victoria, B.C., Canada
    \and
    S.C. Somosan\\
    NewHeights Software Corporation\\
    Victoria, B.C., Canada
}

\maketitle
\thispagestyle{empty}

\begin{abstract}
For the right application, the use of programming paradigms such as
functional or logic programming can enormously increase productivity
in software development. But these powerful paradigms are tied to exotic
programming languages, while the management of software development
dictates standardization on a single language.

This dilemma can be resolved by using object-oriented programming in a
new way. It is conventional to analyze an \emph{application} by
object-oriented modeling. In the new approach, the analysis identifies
the paradigm that is ideal for the application; development starts with
object-oriented modeling of the \emph{paradigm}.  In this paper we
illustrate the new approach by giving examples of object-oriented
modeling of dataflow and constraint programming.  These examples
suggest that it is no longer necessary to embody a programming paradigm
in a language dedicated to it.
\end{abstract}

\Section{Introduction}

What programming language should we use? The answer to this question
has changed over the decades. In the 1970s the answer was: ``The Right
One''. Since then it has become: ``What Everyone Else Is Using''.  For
example, in the 1970s one company embarked on the design and
implementation of a language that was to be ideal for developing
telephone switch software. Although they were successful, in the 1990s
they judged it more important to use a standardized language with
multiple and competing vendors. Accordingly, the ideal language was replaced by
{\tt C} and {\tt C++}, much to the detriment of their subsequent software
development.

Not only in this company, but in almost every other organization, a
similar shift has occurred.  In this paper, we want to re-examine the
now discarded answer, ``The Right One.'' Why was this ever considered
the right answer?

We believe it was based on the observation that with
the languages such as Prolog, Scheme or ML, some problems become
miraculously easy to program. But that depends on the problem: if it is
easy in Scheme or ML, it may not be so in Prolog, and vice versa. Thus, the
effect depends on the \emph{programming paradigm} on which the language
is based: functional programming in the case of Scheme and ML; logic
programming in the case of Prolog.

The fact that, with the right choice of language some problems become
miraculously easy to program we call the \emph{Whitehead effect},
inspired by the following quote from
Alfred North Whitehead (1861-1947):
\begin{quote}
``By relieving the brain of unnecessary work, a good notation sets it free to
concentrate on more advanced problems, $\ldots$ ''
\end{quote}
Whitehead goes on to claim that the effect is to increase the mental capacity
of those who use the notation.

In this connection one should also note the \emph{Sapir-Whorf
hypothesis}, stating that what one can think is determined by one's
language.  In this form the hypothesis is vague in the extreme. Though
various attempts at making it concrete have been discredited, the
hypothesis may have something to do with the fact that, with the right
language, some things become miraculously easy to program.

\SubSection{Radical Software Development}
The Whitehead effect suggests that one first determines the
programming paradigm that makes the application miraculously easy to
program and that one then uses a language that embodies the paradigm.
This we call \emph{Radical Software Development}.

It is easy to see why this approach is not practical: 
Radical Software Development tends to lead to different paradigms, as
required by the different applications. As it is, paradigms are locked up in
programming languages. As a result Radical Software Development
requires writing modules in several programming languages.

\SubSection{A One-Language World}
Meanwhile, in the real world, companies are increasingly under pressure
to ``stick to their knitting''. For example, this means that a company
making telephone switches gets out of developing and maintaining its
own ideal programming language.  Managers want to cut costs of training
new hires and insist on a widely used, standardized language. For
further economies, such a language should come with tools and multiple,
competing vendors. This explains why the world of programming has
become a one-language world, or at least has been trying to.

It is not just managers who are opposed to multiple programming
languages. The Right Language tends to be an exotic one, and does not
sit well on the resume of the programmer who may soon need to look for
another job.  Another problem with The Right Language is what we call
the \emph{Ninety-Percent Phenomenon}, which we will now explain by an
example.

Suppose one finds that logic programming is the paradigm that makes the
application miraculously easy. At present, this means using Prolog. A
browse of the manual shows that ninety-percent of it is taken up by
matters that have nothing to do with \emph{logic} programming. This
ninety-percent is taken up with the mundane infrastructure that all
languages seem to require, regardless of paradigm:  what characters are
allowed in names, what kind of numbers there are, how you write them,
strings, characters, I/O, and so on.  To their credit, the designers of
{\tt C++} and Java have been careful to do this as much as possible as
it has been done in {\tt C}.  In Prolog it has been done independently
of {\tt C}, or of any other language. As a result, Prolog does it
sometimes the way you guess it will do it, and sometimes not. Rather
time consuming, and frustrating.

\Section{OO modelling of the paradigm}
In object-oriented programming (OOP) it is routine to model an
\emph{application} from the ground up in terms of objects.
In one widely used approach \cite{wbww90}, one notes the nouns and
the verbs of a description of the application in English. The nouns are
then considered as candidates for the classes, the verbs as candidates
for the methods.
If one can do this for transactions, invoices, customers,
$\ldots$, why not for the key concepts that make a particular
programming paradigm miraculously effective for the application?
In this
paper we argue that OOP can be used to do this for programming
\emph{paradigms}.

Let us consider functional programming.  In procedure-ori\-ent\-ed
languages, numbers are privileged in that one can (1) give them names,
(2) assign them to variables, and (3) return them as function results.
In these languages, functions are underprivileged in that they share
with numbers property (1), but not (2) and (3). The motivation for
functional programming languages is to accord to them all the
privileges of numbers, and thus make them ``first-class objects'', as
was the going terminology in functional programming \cite{sto77}.

Thus it was apparently common to regard functions as objects before
1977. It was only a matter of time to work out the consequences of this
view. In fact, it happened twenty years later in Friedman and
Felleisen's \emph{A Little Java, A Few Patterns} \cite{frfl97}.
However, it took the form of a cryptic one-page appendix.

From this page, it is apparent that the absence of generics in Java
makes the modelling of higher-order functional programming a tiresome
exercise.  To overcome this limitation, one could have followed {\tt
C++} and added templates. However, Wadler and Odersky judged the
Hindley-Milner type system, used in ML, as a superior alternative. They
extended Java to include this type system. This resulted languages such as Pizza
and GJava
\cite{drskwdlr97,bosw98} that directly allow higher-order functional
programming, without the need for object-oriented modelling in the
sense of this paper.
The Java extended in this way became the standard in 2004 under the name
``Java 2 Platform, Standard Edition 5''.

Pizza and GJava are examples of Multi-Paradigm Object Oriented
Programming Languages. Here the object-orienta\-tion of Java is used to
extend it to include another programming paradigm. Several other
projects in this direction were reported at MPOOL 2001 \cite{dvsmst01},
a workshop in conjunction with the ECOOP conference.

As we explained, there is a strong tendency for one language to
dominate practical programming. Although a multi-para\-digm OO language
can be more elegant and powerful, it is unlikely to be accepted in
practice in the foreseeable future.  It is therefore of interest to see
how far one can go in the direction of multiparadigm programming within
a programming language that is widely used in practice. In this paper
we show by worked examples in {\tt C++} that multi-paradigm
programming is not only possible, but can be quite simple and elegant.
To fit the paper's format, we need to restrict ourselves to
small paradigms.  As we will see, dataflow and constraints are
small enough to exhibit as worked examples here.

\Section{Dataflow programming}

In the dataflow paradigm all computation happens in a network
consisting of nodes connected by unidirectional datapipes. Thus each
node has zero or more input pipes (when the output end of the pipe is
connected to the node), and zero or more output pipes (when the input
end of the pipe is connected to the node). A pipe is used by repeatedly
placing data items at its input end.  These items can be retrieved from
its output end in the same order in which they entered at the opposite
end.  Abstractly viewed, the pipes behave like the abstract data
structure referred to as a {\em queue}.

The dataflow paradigm is justified by the class of problems that it
makes easy to solve.  For example, such problems arise in business
process re-engineering.  Such processes are naturally analyzed and
re-designed in terms of workflow \cite{jcktwd97}.  The automation of
workflow then naturally translates to dataflow.

An older method is Structured Systems Analysis (also called SADT, for
Structured Analysis and Design Technique), which was at one time widely
applied in commercial dataprocessing \cite{ganeSarson79}. However,
Structured Systems Analysis was a world unto itself, apparently not
aware of the larger context of dataflow programming.  However,
Structured Systems Analysis was a world unto itself, apparently not
aware of the larger context of dataflow programming.

Structured Systems Analysis discovered dataflow by inspired thinking
about commercial dataprocessing applications. Ashcroft and Wadge
\cite{ashcrWadge85} arrived at the idea starting from studies in
semantics of programming languages. Dennis, Arvind and others at MIT
have arrived at the dataflow paradigm from computer architecture
\cite{veen86}.  Dataflow has been identified as a Software Architecture
\cite{shwgrln96}.

As early as 1977 the paradigm was already sufficiently compelling that
a programming language was designed to make dataflow programming as
natural as possible \cite{kahnmcq77}.  The paper just mentioned also
contains some simple and widely appealing examples showing the paradigm
at its best.  Note that the paradigm was independently arrived at from
disparate areas: business applications, semantics, and architecture
(hardware and software).  This suggests that it's ``real'' in some
sense.

\SubSection{Hamming's problem solved in dataflow}
A good introduction to the dataflow paradigm is {\em Hamming's problem}
\cite{kahnmcq77}:
\begin{quote}
to print out in increasing order
all positive integers that have no prime factors other than
2, 3, or 5.
\end{quote}
Thus, the sequence starts with 1, 2, 3, 4, 5, 6, 8, 9, 10, 12, 15,
$\ldots$
This problem is attributed to R.W. Hamming by Dijkstra \cite{djk76},
who provided an ingenious solution. It is more efficient than, but is
not as easy to understand as, the dataflow version given by
Kahn and McQueen \cite{kahnmcq77}, which we follow here.

One approach to a solution starts with the observation that the infinite
sequence $x$ of numbers required by Hamming's problem satisfies the
following equation:
$$ x = 1 \circ \mr(\mr(t_2(x),t_3(x)),t_5(x)) $$
where
\begin{itemize}
\item
the meaning of $\circ$ is defined by
$u \circ v$ being the result of prefixing the sequence $v$ by the
element $u$
\item
$\mr$ is the result of merging its two sorted input sequences
into a sorted output sequence, suppressing duplicates
\item
$t_2$ is result of multiplying by 2 its input sequence, element by
element; similarly for $t_3$ and $t_5$
\end{itemize}

The question whether the solution to Hamming's problem is the {\em only}
solution to the equation is addressed by the methods developed in Kahn
\cite{kahn74}.

To turn this observation into a dataflow network, we take the above
equation with a complex expression and turn it into a system of simple
equations by introducing auxiliary variables. We do this in two steps.
In the first step, we get rid of nested expressions:

\begin{eqnarray*}
&&a = t_2(x_2) \;\;\;\;       b   = t_3(x_2) \;\;\;\;   d = t_5(x_2)   \\
&&c = \mr(a,b)  \;\;\;\;  x_1 = \mr(c,d)               \\
&&x_2 = 1 \circ x_1 
\end{eqnarray*}

The resulting equations can, if considered in isolation, each be
translated directly into a node of a dataflow network.  Each of $a$,
$b$, $c$, $d$, and $x_1$ correspond to a datapipe because, in the above
set of equations, they have exactly one occurrence in a left-hand side
and exactly one occurrence in a right-hand side.  An occurrence on the
left-hand (right-hand) side corresponds to the output (input) side of
the datapipe.

However, $x_2$ has too many occurrences in right-hand sides.  We can
avoid this problem by making these occurrences into different
variables, say $f$, $g$, and $h$. But how do we tell that these are the
same sequence?  To do that, we introduce a node type, with one input
and two output pipes, that outputs two identical copies of each item
that it removes from the input pipe.  Let us call this node $\spl$.

\begin{eqnarray*}
&&i = x_2  \;\;\;\;    h = x_2\;\;\;\;   f = i\;\;\;\;   g = i      \\
&&a = t_2(f) \;\;\;\;      b = t_3(g)\;\;\;\;       d = t_5(h)   \\
&&c = \mr(a,b) \;\;\;\;    x_1 = \mr(c,d)                \\
&&x_2 = 1 \circ x_1                  
\end{eqnarray*}

\begin{figure}[htbp]
\begin{center}
\epsfig{file=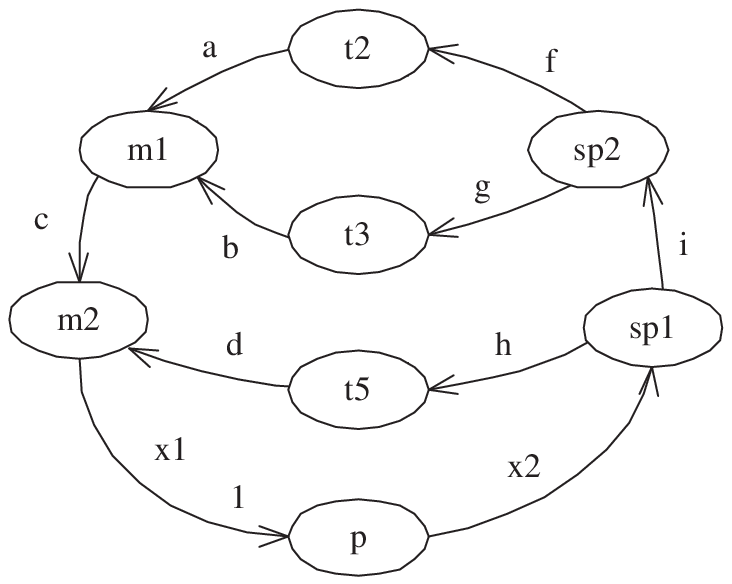}
\end{center}
\caption{\label{fig:network} Dataflow network for Hamming's problem in 
the initial state, where all pipes are empty except a 1 in {\tt x1}.}
\end{figure}

The entire top two lines each correspond each to a $\spl$ node;
the remaining six equations correspond to a node each. This makes eight
nodes in all.  See the network diagram in Figure \ref{fig:network}.

\SubSection{An object-oriented implementation of Hamming's problem}
The most widely known principle of object-oriented modelling is to
consider the nouns of the specification as candidates for classes.

In any dataflow network, particularly conspicuous nouns are {\em node}
and {\em datapipe}.  As observed before, datapipes behave like queues,
a commonly used class.

According to the principle just mentioned, we should consider a class
suitable for creating all required nodes as instances.  The principle,
though a good first approximation, needs some refinement. This is
because objects of the same class should have states of the same form,
though not necessarily of the same content. The state of a node object
includes the states of the abutting datapipes. And of course, instances
of the same class should have the same behaviour.

These considerations suggest that the $\mr$ nodes are instances of the
same class ({\tt merge}; the individual instances are {\tt m1} and {\tt
m2}), as are the $\spl$ nodes (of the class {\tt split}, with instances
{\tt sp1} and {\tt sp2}), as are the nodes $t_2$, $t_3$, and $t_5$ (of
the class {\tt times}, with instances {\tt t1}, {\tt t2}, and {\tt
t3}).

\begin{figure}[htbp]
\begin{verbatim}
01: int main() {
02:   const int MaxTimes = 50;
03:   queue a(10), b(10), c(10), d(10),
04:         x1(10), x2(10),
05:         f(10), g(10), h(10), i(10);
06:   merge m1(&a,&b,&c), m2(&c,&d,&x1);
07:   times t2(2,&f,&a), t3(3,&g,&b),
08:         t5(5,&h,&d);
09:   split sp1(&x2,&h,&i), sp2(&i,&f,&g);
10:   print p(&x1,&x2);
11:
12:   node* ar[] = { &m1, &m2, &t2, &t3,
13:	        &t5, &sp1, &sp2, &p };
14:   int arSz = sizeof(ar) / sizeof(node*); 
15:    
16:   x1.add(1);
17:   for ( int i=0; i < MaxTimes; i++ )
18:       for ( int j=0; j < arSz; j++ )
19:           ar[j]->run();
20:   return 0;
21: }
\end{verbatim}
\caption{\label{fig:dataFlowProg} {\tt C++} code for the dataflow network
for Hamming's problem. The node {\tt print} has been added to allow the 
solution to be printed.}
\end{figure}

Lines 3--5 create instances of class {\em queue} to act as datapipes,
each with a maximum size arbitrarily set at 10.  Given these pipes,
lines 6--10 create the nodes, with the correct pipe connections.
\emph{These lines seem the most succinct possible textual
representation of the diagram in Figure~\ref{fig:network}.  In so far
as this is true, $\cpp$ comes close to the best possible dataflow
programming language.}

Lines 3--10 {\em create} the dataflow network; they do not cause it to
execute its computation. The attraction of the dataflow paradigm is
to avoid the difficulty of conventional programming, namely to ensure 
that events happen in the right sequence. To execute a dataflow
network, each node executes, independently of the others, the following
simple computation:
\begin{quote}
\emph{If any of the input datapipes is empty, or if any of
the output datapipes is full, do nothing. Otherwise, remove the next
item from each of the input pipes, perform on them the specialized
computation characteristic for the type of node, and place the
results, if any, on the output pipes.
}
\end{quote}
The computation just described is invoked by a method called {\tt run},
which is defined for each of the classes {\tt merge}, {\tt times}, and
{\tt split}.

To execute the entire network, one invokes the {\tt run} method for
each node, as in lines 18--19 of Figure~\ref{fig:dataFlowProg}. 
Typically several of these
invocation have no effect because of full output or empty input pipes.
But if the network can do anything at all, then at least one node will
do something. In large networks it is worth optimizing the invocations
of the {\tt run} method. One can keep track of which nodes are
blocked.  A blocked node connected to a pipe of which the content
changed may no longer be blocked and becomes a candidate for being {\tt
run}. Such an optimization is reminiscent of the constraint propagation
algorithm of D. Waltz \cite{wltz75}.

Note that in lines 12--13 the nodes of the dataflow are placed into an
array. The order of the nodes in this array gives the order in which the 
{\tt run} methods are called (line 19). However, this order does not 
matter, since in dataflow the order in which things are done matters less
than it does in conventional programming.

Space limitations prevent us from listing the entire program, which is
about a hundred lines, including the queue implementation. We just 
add some representative code:
\begin{verbatim}
class node {
public:
  virtual void run() = 0;
};

class times : node {
private:
 int mult; queue *in, *out; 
public:
 times(int Mult, queue *In, 
       queue *Out) {
   mult = Mult; 
   in = In; 
   out = Out;
  }
  void run() {
   if (in->empty() 
          || out->full()) 
      return;
   out->add(mult*(in->next()));
   in->remove();
  }
};
\end{verbatim}
An attractive characteristic of dataflow is that the nodes can run
concurrently subject to mutual exclusion on the datapipes.
We have considered doing this example in Java to make it easy to have
every instance of {\tt node} run in its own thread.
But although threads are simple to use, the result is still not as simple
as doing without. To find Hamming numbers, threads are not essential.
So we make our point better by showing a simpler program that just gets
the numbers.

\Section{Constraint programming}

Many problems in resource planning and numerical computation can be
solved in a \emph{declarative} way: one states the relations that are
to hold between the unknowns; a suitable solver then finds values such
that all relations are satisfied. The relations are referred to as
\emph{constraints}\/; the method is known as \emph{constraint
programming} \cite{mrrst98,jln01}.

Constraint programming is useful because systems can be solved that
contain thousands of constraints. Here we will of course illustrate with
a very small example.

\SubSection{Complex constraints}
A particular constraint programming method is \emph{interval
constraints}.  Specific for interval constraints is that the unknowns
are real numbers and that their domains are intervals.  As an example
consider the problem of finding the intersection points of the
circle $x^2+y^2=1$ and the parabola $y=x^2$. This means finding values
for $x$ and $y$ such that both relations are satisfied.

Of course a student in secondary school will identify $y$ as
$(\surd 5 -1)/2$, and $x$ accordingly.
The point here is to develop a system that determines solutions
directly from the set of constraints as given,
for a wide class of constraints.

Suppose that
initially all we know is that $x$ and $y$ are in $[-\infty,+\infty]$.
Considering each of the two given relations separately would already
remove some values for $x$ or $y$ from consideration. For example, it
is clear from the constraint $x^2+y^2=1$ that $x$ and $y$ have to be in
$[-1,+1]$.  Whatever it is that allows us to reduce the original
intervals $[-\infty,+\infty]$ to $[-1,+1]$ we call a \emph{contraction
operator}. It is an operator associated with the constraint that allows
one to make such an inference.  In constraint programming, one applies
in turn the contraction operators associated with the constraints until
no more contraction results. The remaining intervals for the variables
then give all information about the solutions of the problem that this
method can give.

To make this method practical, contraction operators have to be widely
applicable and efficiently implementable. Such operators have only been
discovered for a relatively small repertoire of \emph{primitive}
constraints. These include constraints with binary relations such as $x
\leq y$, $x = y$, and $ y = x^2$.  There are also constraints with
ternary relations such as $x+y=z$ and $x \times y = z$.  The primitive
constraints do not include the constraint $x^2+y^2=1$, as it does not
occur sufficiently often to justify its presence in a general-purpose
solver.  This complex constraint is therefore decomposed into primitive
constraints with the aid of auxiliary variables. In this way the
circle-and-parabola problem is expressed by the system \begin{equation}
\label{eq:diss} x^2=x_2, \: y^2=y_2, \: x_2+y_2=1, \: x_2=y
\end{equation}

\SubSection{Contraction operators}
To introduce contraction operators, let us look at an example of one in
action: the one for primitive constraints of the form $x+y=z$.  Such a
constraint expresses that the variables are in a ternary relation that
we call \emph{sum}.  If it is known initially that $x$ and $y$ are in
$[0,2]$ and that $z$ is in $[3,5]$, then it is clear that neither $x$
nor $y$ can be near $0$ and that $z$ cannot be near $5$. In fact the
contraction operator reduces the intervals for $x$ and $y$ to $[1,2]$
and the one for $z$ to $[3,4]$.

This is optimal and can be computed in a few floating-point
operations.  One can solve complex systems of nonlinear equations and
inequalities by first reducing them to primitive constraints, and by
then applying contraction operators until nothing changes. Typically,
the remaining intervals are too large to be useful, in which case one
interval is split and the constraint propagation is repeated.

\SubSection{OO modelling of interval constraints}
Let us now do a perfectly straightforward exercise in
object-oriented modelling in the spirit of \cite{wbww90}.
Clearly, ``constraint'' is an important noun, so justifies a class of
that name. The same holds for ``variable''. However, this is a
dangerously ambiguous concept: in programming languages, in logic, and
in calculus it means different things. The role of $x$ in the example
can be stated precisely: it is the name of a real number that we do not
know. Thus $x$ is an instance of a class of which the instances represent
real numbers. This suggests the following class definition:
\begin{verbatim}
class real {
private:
  FLPT lb, ub;
public: 
  real(); 
  real(FLPT lb, FLPT ub);
};
\end{verbatim}
{\tt FLPT} is a generic floating-point number type; it could be of
single or double length.  The class {\tt real} is a classic example of
object-oriented modelling.  It stands for an abstract concept, in this
case a real number. It hides the representation, which is the
description of a set of real numbers to which the real number belongs.
The set is restricted to the form of an interval {\tt [lb,ub]}, which
has the floating-point numbers {\tt lb} and {\tt ub} as lower and upper
bounds.

In the case of real numbers this distinction between the abstract
concept, which is public, and its hidden representation is especially
valuable because most real numbers do not even have a finite
representation. So far, this perfectly ordinary piece of
object-oriented modelling has been, as far as we know, the only
practical way to directly compute with mathematical models involving
continuously varying quantities.

The zero-argument constructor creates a real of which nothing is
known.  Accordingly it is represented by the interval
$[-\infty,+\infty]$, which is the set of all real numbers. If more is
known about the real number, then it is represented by a smaller
interval. In our example we need the real numbers 1 and 0.5, which are
represented by the intervals only containing these numbers and are
created by the constructor calls \verb+real _1(1.0,1.0)+ and
\verb+real _0(0.5,0.5)+.

To model constraints, we note that their indentity is determined both
by the relation and by the variables connected by the relation. Each
instance of a constraint needs to record the variables of that
instance. It is essential that different constraints be able to share
reals. This sharing is modelled by members that are pointers to
instances of class {\tt real}.  The relation is represented by ensuring
that constraints are instances of the same class if, and only if, they
have the same relation. In this way the contraction operator for that
relation can be a method of that class.  This method is named {\tt
shrinc}\footnote{ It is intended to shrink intervals. Moreover, the
class is part of a system called ``Sound High-Resolution Interval
Numeric Calculator''.  }.
As all constraints need to go into the same container, and as they all
have a {\tt shrinc()} method, all constraint classes derive from an
abstract class named {\tt constraint}.  These considerations are
embodied in the code in Figure~\ref{fig:headers}.

\begin{figure}[htb]
\begin{verbatim}
class constraint {
public:
  virtual bool shrinc() = 0;
  virtual ~constraint {}
  // Applies contraction operator.
  // Returns false iff 
  //  an empty interval results.
};
// constraint is  x <= y
class leq: public constraint {
  real *x, *y;
public:
  leq(real *x, real *y);
  bool shrinc();
};
// constraint is  x == y
class eq: public constraint {
  real *x, *y;
public:
  eq(real *x, real *y);
  bool shrinc();
};
// constraint is  x+y = z
class sum: public constraint {
  real *x, *y, *z;
public:
  sum(real *x, real *y, real *z);
  bool shrinc();
};
// constraint x^2=y
class square: public constraint {
  real *x, *y;
public:
  square(real *x, real *y);
  bool shrinc();
};
\end{verbatim}
\caption{\label{fig:headers}
Definitions for the {\tt constraint} classes.
}
\end{figure}

Now that we have the interface with the interval constraint machinery,
let us use it to express the system (\ref{eq:diss}) for the
circle-and-parabola problem; see Figure~\ref{fig:propCode}.

\begin{figure}[htb]
\begin{verbatim}
00: int main () {
01:  //solve {x^2=x2, y^2=y2, x2+y2=1, y=x2}
02:  //create variables:
03:  real x, y, x2, y2, _1(1.0,1.0);
04:  //real _0(0.5,0.5);
05: 
06:  //create constraint system:
07:  constraint* array[] = {
08:    new square(&x,&x2), new square(&y,&y2),
09:    new sum(&x2,&y2,&_1), new eq(&y,&x2)
10:    //, new leq(&_0,&x)
11:   };
12:  //propagate:
13:  const int MAX = 1000;
14:  int size =
      sizeof(array)/sizeof(constraint*);
15:  for (int i=0; i < MAX; i++)
16:    for (int j=0; j < size; j++)
17:      array[j].shrinc();
18:  cout << "x: " << x << endl;
19:  cout << "y: " << y << endl;
20:  for (int j=0; j < size; j++)
21:    delete array[j];
22: }
\end{verbatim}
\caption{ \label{fig:propCode}
$\cpp$ code for finding the intersection of a circle with a parabola
using the interval constraints method.}
\end{figure}

In line 3 the variables of system (\ref{eq:diss}) are created.
Lines 7 through 11 create the constraints of this system
and place them in a container. At the same time, this specifies
the interconnections of the constraints via the shared variables.

After the constraint system has been built, contraction operators are
applied. This is done in a simplistic way in lines 13 through
18.  In actual practice, it is done by means of a constraint
propagation algorithm. These algorithms have a long and interesting
history; see \cite{aptEssence} for recent contributions and a history.

The result of the propagation is printed on lines 18
and 19 with the result: $x \in [-1,1]$ and $y \in [0,1]$.
This result is large enough to contain both
solutions to (\ref{eq:diss}). If one is interested in one solution,
say, with
positive $x$, then one can add a constraint as in line 10, using a
constant created as in line 4.
In that case there is one solution, enclosed in a narrow interval:
\begin{eqnarray*}
x \in [\overline{0.78615137775742}29,\overline{0.78615137775742}36] \\
y \in [\overline{0.61803398874989}44,\overline{0.61803398874989}54].
\end{eqnarray*}

We include this example because we believe
that it shows that a special-purpose constraint programming language
would not be able to specify the system (\ref{eq:diss})
in a clearer or more succinct way.
At the same time,
a special-purpose language
would probably not have any original insights
on how to do containers, nor on how to do control primitives,
so that these aspects of the program would be the same
or gratuitously different.
Hence, this straightforward exercise
in object-oriented modelling of constraint programming
is probably as well as one can do in the form of text.

\Section{Object-oriented frameworks}

The foregoing examples suggest that a distinctive programming paradigm
need not have a language of its own. It is adequately supported
by a suitable class library.
Both our examples include an abstract class and a generic algorithm
expressed in terms of that class. Programming in the paradigm requires
one to define derived class suitable to the application. In our
dataflow example, each type of node and its specific activity becomes
a class derived from {\tt node}.
Similarly, in the constraint example, each type of constraint and its
specific contraction operator becomes a class derived from {\tt
constraint}.

But the library, with its abstract classes and generic algorithms is
not enough. If it indeed embodies a distinctive programming paradigm,
it comes with a view of how to do things, a mindset. This can be
expressed by a set of tutorial examples. The collection of these things
is what is called a \emph{framework} by Gamma, Helm, Johnson, and
Vlissides \cite{ghjv94}.

The trend towards frameworks rather than dedicated programming languages
may have started with Puget's approach to constraint programming.
At first, constraint programming followed the conventional route with a
dedicated programming language. It grew out of logic programming, which
had Prolog as dedicated programming language.
Accordingly, constraint programming was implemented in various dialects
of Prolog: Prolog II and III, CHIP and its descendants, BNR Prolog,
and Prolog IV. However, Puget adopted Saraswat's comprehensive view of
constraint programming \cite{srswt93} called
\emph{Concurrent Constraint Programming}, dropped the concurrency,
and based a {\tt C++} class library on it \cite{pgt94,pgtlcnt95}.

Design Patterns and object-oriented frameworks seemd equally promising when these
ideas first arose.
Design patterns found wide acceptance. However, to quote Erich Gamma
\cite{gmm02}
\begin{quote}
When we wrote ``Design Patterns''
we were excited about frameworks
and forecast that there would be lots of them in the near future.
Time has shown, however, that frameworks aren't for everyone.
Developing a good framework is hard
and requires a large up-front investment that doesn't always pay off.
\end{quote}
Originally, frameworks were intended to help specific applications,
such as graphical user interfaces or document processing.
Like design patterns, object-oriented frameworks
are defined as systems of customizable co-operating classes.
What one obtains as a result of modelling a programming paradigm
also answers to this description, but is not a design pattern.
Instead, it is useful as a new way of discovering
useful systems of customizable co-operating classes.

One of the most important advantages of design patterns
is as an aid to documentation:
just because the pattern has a name,
the use of this name in the documentation
speeds up understanding of the part of the code concerned.
The object-oriented frameworks arising from programming paradigms
have this same advantage. 

\Section{Conclusions}

So far there has been little scope in practice for the Whitehead
effect to ease software development. Applying the paradigm that's
right for the application seemed to require switching to a different
programming language. Practical concerns necessitate sticking to a
single language.

This situation has changed since the single language is
object-oriented. Programming paradigms such as dataflow or
constraints, which once were thought to need a dedicated language, are
easily modelled in an object-oriented language. In fact, as easily as
the textbook example of object-oriented modelling of simple
applications. Thus it is not overly ambitious to model the ideal
programming paradigm, rather than the application.
This modelling typically takes the form of an OO framework.

In this we can distinguish between minor and major paradigms. The minor
ones are those that can be modelled in one of the few languages most
widely accepted in industry. We saw that dataflow and constraint
programming fit in this category.  J\"arvi and Powell \cite{jrpw01}
implement partial function application in this way in {\tt C++}.  The
fact that functional programming can be characterized by functions
being first-class objects, suggests that functional programming is no
harder to do. It has turned out that it is and that it seems to require
a new programming language.  Apparently, functional programming is a
major paradigm. However, even in this case OO programming has made a
difference. The new language turned out to be an extension of Java that
was needed anyway, independently of functional programming.

The question remains whether logic programming, so far sequestered in
Prolog and other dedicated languages, is a major paradigm requiring a
language extension or a minor one that can easily be modelled in an
existing OO language. The work of combining Smalltalk with logic
programming by symbiotic reflection \cite{wtsdcs01} seems to transcend the
simplistic criterion used to distinguish between minor and major
paradigms.

It used to be that not only paradigms, but certain applications got
their own programming languages. We have argued that this is something
of the past.  It was even the case that one application, namely
simulation, had several languages devoted to it. One of these was
Simula, and the rest is history.


\end{document}